\documentclass[%
 aip,
jap,
 sd,%
 amsmath,amssymb,
 reprint,%
]{revtex4-1}

\usepackage{graphicx}
\usepackage{dcolumn}
\usepackage{bm}

\begin{document}

\preprint{AIP/123-QED}

\title[A method for controlling the magnetic field near a superconducting boundary in the ARIADNE axion experiment]{A method for controlling the magnetic field near a superconducting boundary in the ARIADNE axion experiment}

\author{H. Fosbinder-Elkins}
 \email{harryfe@nevada.unr.edu}
 \altaffiliation{Physics Department, Princeton University}
 \affiliation{Department of Physics, University of Nevada, 1664 N Virgina St., Reno, NV, 89557}
\author{J. Dargert}
 \affiliation{Department of Physics, University of Nevada, 1664 N Virgina St., Reno, NV, 89557}
 \author{M. Harkness} 
 \affiliation{Department of Physics, University of Nevada, 1664 N Virgina St., Reno, NV, 89557}
\author{A.A. Geraci}
\email{andrew.geraci@northwestern.edu}
 \altaffiliation{Department of Physics and Astronomy, Northwestern University, Evanston, IL}
  \affiliation{Department of Physics, University of Nevada, 1664 N Virgina St., Reno, NV, 89557}
 \author{E. Levenson-Falk}
  \affiliation{Physics Department, Stanford University, 382 Via Pueblo, Stanford, CA 94305}
 \author{S. Mumford}
  \affiliation{Physics Department, Stanford University, 382 Via Pueblo, Stanford, CA 94305}
 \author{A. Kapitulnik}
  \affiliation{Department of Physics and Applied Physics, Stanford University, 382 Via Pueblo, Stanford, CA 94305}
 \author{Y. Shin}
  \affiliation{IBS Center for Axion and Precision Physics Research, KAIST, 193 Munji-ro, Yuseong-gu, Daejeon 34051, South Korea}
 \author{Y. Semertzidis}
   \affiliation{IBS Center for Axion and Precision Physics Research, KAIST, 193 Munji-ro, Yuseong-gu, Daejeon 34051, South Korea}
 \author{Y.-H. Lee}
\affiliation{KRISS, 267 Gajeong-ro, Yuseong-gu, Daejeon 34113, Republic of Korea}
\date{\today}

\begin{abstract}
The QCD Axion is a particle postulated to exist since the 1970s to explain the Strong-CP problem in particle physics. It could also account for all of the observed Dark Matter in the universe. The Axion Resonant InterAction DetectioN Experiment (ARIADNE) experiment intends to detect the QCD axion by sensing the fictitious ``magnetic field'' created by its coupling to spin. The experiment must be sensitive to magnetic fields below the $10^{-19}$ T level to achieve its design sensitivity, necessitating tight control of the experiment's magnetic environment. We describe a method for controlling three aspects of that environment which would otherwise limit the experimental sensitivity. Firstly, a system of superconducting magnetic shielding is described to screen ordinary magnetic noise from the sample volume at the $10^8$ level. Secondly, a method for reducing magnetic field gradients within the sample up to $10^2$ times is described, using a simple and cost-effective design geometry. Thirdly, a novel coil design is introduced which allows the generation of fields similar to those produced by Helmholtz coils in regions directly abutting superconducting boundaries. The methods may be generally useful for magnetic field control near superconducting boundaries in other experiments where similar considerations apply.
\end{abstract}

\maketitle

\section{Introduction}
The Axion Resonant InterAction Detection Experiment (ARIADNE) intends to search for the QCD axion using techniques based on nuclear magnetic resonance \cite{arv14,ger17}.  Axions or axion-like particles will generically mediate short-range spin-dependent interactions between an ensemble of nuclear spins and an (un-polarized) attractor mass.  In the experiment, a sample of laser-polarized $^3$He nuclear spins feels a fictitious “magnetic field” as the teeth of a sprocket-shaped tungsten attractor rotate past the sample at the nuclear Larmor precession frequency, set to be approximately 100 Hz \cite{ger17}. Here the axion is acting as the mediating boson responsible for the interaction.  The experiment has the potential to probe deep within the theoretically interesting regime for the QCD axion in the mass range of 0.01-10 meV, while being independent of cosmological assumptions. Detecting the axion would explain the smallness of $\theta_{\rm{QCD}}$ \cite{axion1, axion2,PTViolation, Moody:1984ba} and identify a component of Dark Matter \cite{arv14,PTViolation,ADMX}. 

The axion can mediate an interaction between fermions (e.g. nucleons) with a potential given by \begin{equation}
 U_{sp}(r)=\frac{\hbar^2 g_s^N g_p^N}{8 \pi m_f}\left( \frac{1}{r \lambda_a}+\frac{1}{r^2}\right) e^{-\frac{r}{\lambda_a}} \left(\hat \sigma \cdot \hat r \right),
 \end{equation}
 where $m_f$ is their mass, $\hat{\sigma}$ is the Pauli spin matrix, $\vec{r}$ is the vector between them, and $\lambda_a = h/m_A c$ is the axion Compton wavelength \cite{Moody:1984ba,arv14}, which determines the range over which the interaction extends. The range of interaction can be as small as $\lambda=30$ $\mu$m depending on the mass of the axion \cite{pdg}. For the QCD axion the scalar and dipole coupling constants $g_s^N$ and $g_p^N$ are directly correlated to the axion mass. Since the axion couples to $\hat{\sigma}$ which is proportional to the magnetic moment $\vec{\mu}_N$ of the nucleus, the axion coupling can be treated as a fictitious “magnetic field” $B_{\rm{eff}}$ through an interaction potential 
 \begin{equation} 
 U_{sp}(r)=-\vec{\mu}_N \cdot \vec{B}_{\rm{eff}}.\label{gsgp}
 \end{equation}
In ARIADNE, the fictitious field is generated by a tungsten sprocket. As the teeth of the sprocket rotate past the $^3$He spin sample, the distance between the tungsten material and $^3$He spin sample is modulated, thus resulting in a modulation in the field strength seen by the spins. 
 
 It is vital to note that such a magnetic field as described in Eq. (\ref{gsgp}) does not conform to Maxwell's equations and so cannot be detected directly with a magnetometer such as a superconducting quantum interference device (SQUID)\cite{arv14}. Instead, the nuclear spins in a hyperpolarized sample of $^3$He do couple to the locally-sourced fictitious field, and can precess resonantly with the modulation as in a nuclear magnetic resonance (NMR) experiment. This precession allows the fictitious field to be indirectly detected by using a SQUID magnetometer to measure the real magnetic field generated by the precessing $^3$He nuclear spins\cite{ger17}. The setup relies on superconducting magnetic shielding, required to screen the $^3$He sample from ordinary magnetic noise.  

\begin{figure}[!t]
\begin{center}
\includegraphics[width=0.8\columnwidth]{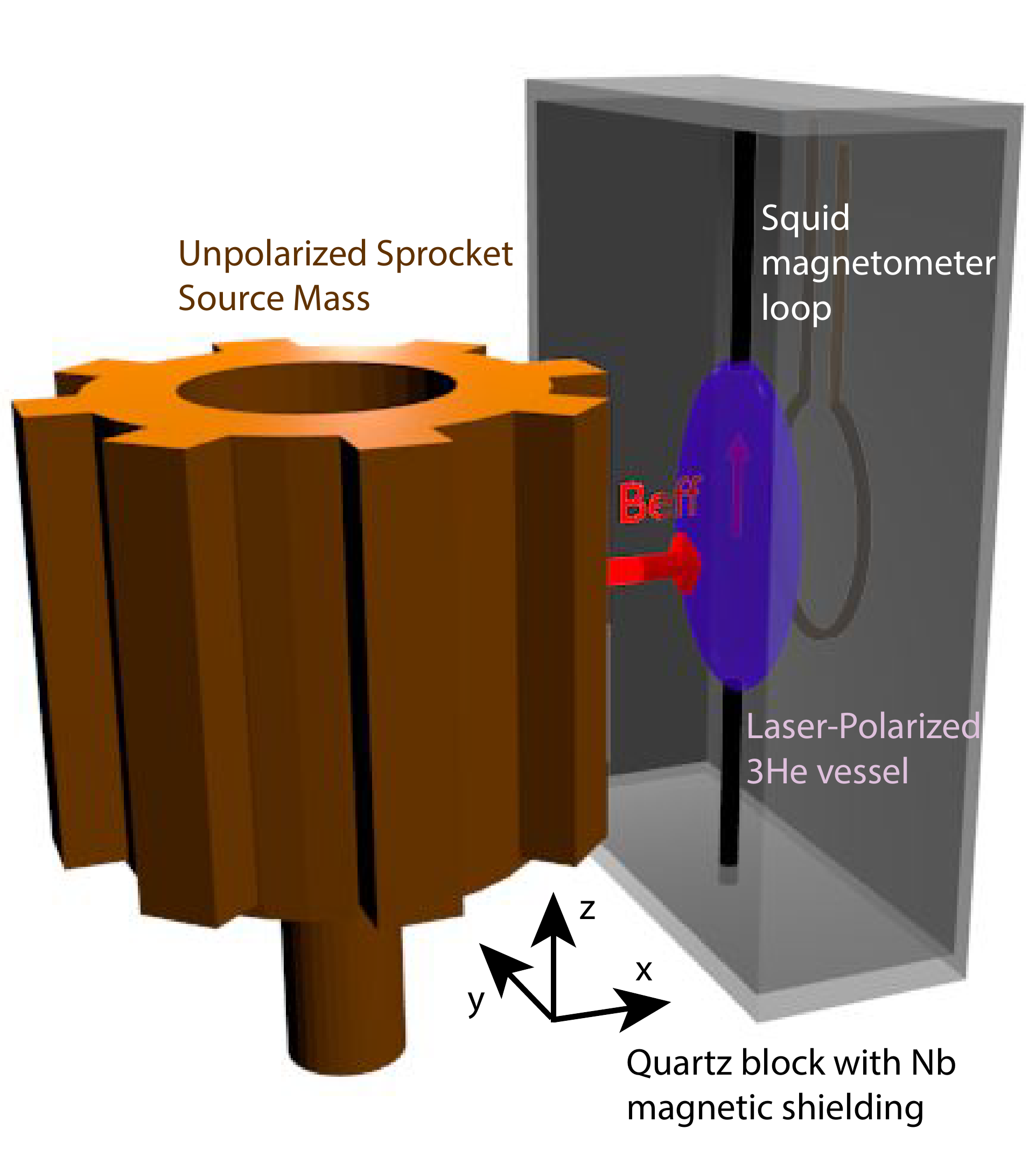}
\caption{Setup: a sprocket-shaped source mass is rotated so its ``teeth'' pass near a laser-polarized $^3$He NMR sample at its resonant frequency, producing a time-varying ``fictitious'' magnetic field via the axion potential. The resulting induced transverse magnetization is read out using a SQUID magnetometer. This distance between the rotating sprocket and He sample is kept within $\approx 200$ $\mu$m to search for short-range axion-mediated forces.  The sample chamber is embedded in a quartz block coated with thin-film superconducting Nb shielding.\label{setup}}
\end{center}
\end{figure}

For the geometry described in \citet{ger17} and illustrated in Fig. \ref{setup}, the fictitious magnetic field felt at the sample is constrained to be very small, less than $10^{-19}$ T. Thus experimental sensitivity is paramount. Under ideal operating conditions the sensitivity of the experiment will be limited by quantum projection noise in the sample itself \cite{arv14}.  The sensitivity is thus maximized for a longer transverse decoherence time $T_2$ of the sample, with the minimum detectable magnetic field scaling as $(n V T_2)^{-1/2}$ where $n$ is the density of polarized spins in a sample of volume $V$. In practice the experimental sensitivity is constrained by several considerations. The first is ordinary (external) magnetic noise, which would exceed the sub-aT axion signal if unmitigated. A second stems from magnetic field gradients in the sample volume, which result in a Larmor frequency which varies throughout the sample, decreasing the effective $T_2$. Thirdly, the sample's nuclear Larmor precession frequency (set by the overall magnetic field felt by the sample) and the modulation frequency of the axion field set by the sprocket's rotational frequency must be matched in an experimentally practical way.

The methods presented in this paper may be generally useful for magnetic field control near superconducting boundaries in other experiments where similar considerations apply, even those relying on detecting the cosmological axion field, such as the Casper Axion Dark matter experiment \cite{casper} or the QUAX proposal \cite{quax}.

 \section{Experimental setup and Statement of Problem}
 
For concreteness, to model the magnetic environment of the magnetized $^3$He sample near a superconducting (SC) boundary, we choose a particular geometry relevant for the ARIADNE experiment. The sample container is chosen to be an oblate spheroid of dimensions $0.15$ mm, $3$ mm, and $3$ mm in the $x$-,$y$-, and $z$- directions, respectively. A pancake-like shape of the sample chamber allows the He detector volume to remain close to the source mass rotor, within the Compton wavelength of the axion. For simplicity, we assume the that the polarization fraction is $1$ and the spins are vertically aligned in the $z$-direction, with a nuclear spin density of $2 \times 10^{21}$ spins/cc, corresponding to the maximal spin density of the $4.2$K $^3$He gas considered in the experiment. The reason for the spheroidal shape is to maintain a uniform magnetic field inside the sample and to maximize the signal by having as much volume of the sample region in close proximity to the source mass, as discussed in detail below.  The spheroid is embedded in a quartz cube of approximate dimensions $5$ cm, coated with a Nb SC film. The quartz thickness between the face of the spheroid and the wall of the cube (in the $x$-direction) is set to be $75$ $\mu$m, sufficient to hold vacuum but thin to allow close proximity to the drive mass sprocket.
  \subsection{Ordinary Magnetic Noise}
At the location of the $^{3}$He sample, ordinary magnetic backgrounds can lie at the $10^{-12} \frac{\rm{T}}{\sqrt{\rm{Hz}}}$ level, well above the expected axion signal, if not screened\cite{ger17}. A primary example of concern is Johnson noise, where thermal motion of electrons in conducting materials produces a spectrum of magnetic field fluctuations near their surface \cite{varpula}. The presence of such noise in fact necessitates the use of superconducting shielding surrounding the sample region, since superconducting materials, being of zero resistance, are immune to such noise.  Other examples of backgrounds include magnetic fields due to the magnetization of the rotating source mass sprocket due to the Barnett effect, and due to the magnetic susceptibility and any magnetic impurities in the sprocket \cite{ger17}. However, because the fictitious magnetic field is not subject to Maxwell's equations, it is not screened by superconducting magnetic shielding.\cite{arv14}. Thus superconducting shielding can be installed to reduce the ordinary magnetic noise felt by the sample without also reducing the fictitious field signal. However, geometry constraints make perfect shielding nontrivial, as wires for sensors must extend into the shielded region. These constraints, and efforts to characterize (and improve) the shielding factor of the resultant design are described in section \ref{design}.  

In order for the experiment to achieve design sensitivity, a shielding factor of approximately $10^8$ is needed at frequencies of $50-100$ Hz \cite{ger17}.  Such shielding factors have been achieved for example with solid Nb superconducting tubes \cite{tubes}. The approach described in this paper requires (at least partial) use of thin film superconducting shielding, due to the requirement of very close proximity (a distance of order $\lambda_a$ or less) between the $^3$He sample and sprocket mass. Experimental work is in progress to evaluate the efficacy of thin film Nb on achieving such shielding factors in the experiment.

  \subsection{Bias field control}
The $^3$He sample is characterized by its Larmor frequency $\omega = \gamma B_0$, where $\gamma$ is the gyromagnetic ratio when the sample is subject to a constant magnetic field $B_0$. During sprocket rotation, the fictitious magnetic field is modulated on resonance with $\omega$ to maximize its effect. To keep the two frequencies as close as possible, both the rotational frequency of the sprocket and the magnetic field felt by the sample ideally are both adjustable. In most cases, Helmholtz coils would be used to create a gradient-free magnetic field to tune the Larmor frequency of the sample. However, the $^3$He sample must remain as close as possible to the rotating sprocket outside the shield, to keep the value of the fictitious magnetic field relatively high. Therefore, the sample cannot sit on the axis of ordinary Helmholtz coils interior to the shield. A design for solving this problem is described in section III.

\subsection{Inhomogeneous Broadening}
\begin{figure}
\includegraphics[width=1.0\columnwidth]{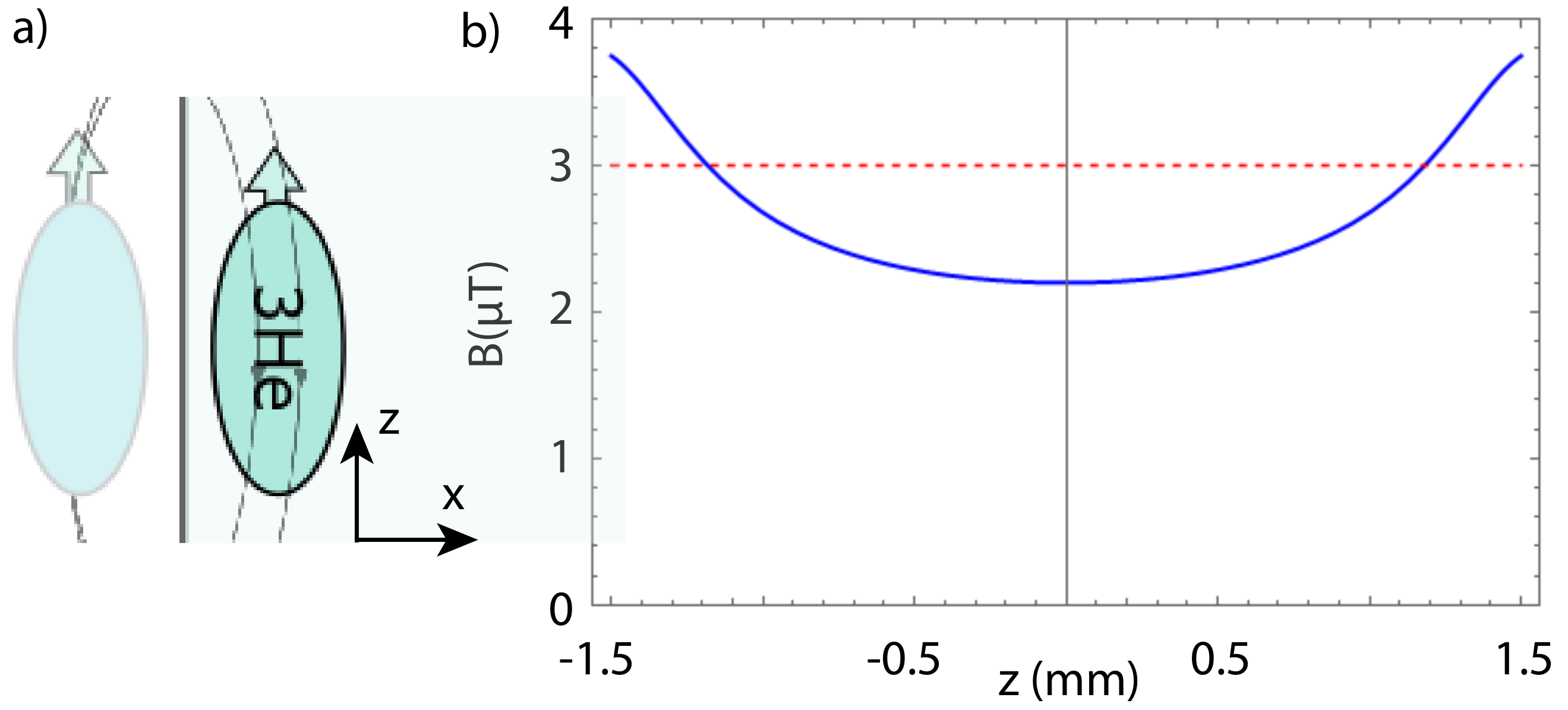}
\caption{\label{image} (a) (not to scale) An illustration of the ``Meissner image'' of the spheroidal sample volume. As for a uniformly magnetized sphere, the magnetic field interior to a magnetized oblate spheroid with the magnetization aligned with one of the principal axes is uniform. When abutting a superconducting surface, the Meissner effect prevents the magnetic flux from penetrating the SC wall.  Mathematically this can be treated with the method of images \cite{methodofimages}. The arrows indicate the magnetizations of the real and image spheroids, while the lines indicate the magnetic field of the image spheroid, which perturbs the otherwise constant field interior to the real spheroid.
(b) Norm of the magnetic field along the z-axis in the spheroid, with a bias field applied of approximately $23$ $\mu$T opposing the direction of magnetization, tuned to achieve an average Larmor frequency around 100 Hz. The dotted red line indicates the field value without the perturbation introduced by the image spheroid. In this configuration the fractional variation due to the perturbation is up to 26\%.}
\end{figure}

If the ordinary magnetic field varies significantly from one part of the $^3$He sample volume to another, it becomes impossible to drive the entire sample on resonance, drastically decreasing the experimental sensitivity and the relaxation time of the $^3$He. To attain maximum sensitivity for the sample geometry and parameters considered, the magnetic field should not vary within the sample by more than $10^{-11} (\frac{1000\rm{ s}}{T_2})$ T, where $T_2$ is the transverse relaxation time.\cite{ger17}. 

Perturbations to the otherwise constant magnetic field internal to the spheroidal sample volume stem from multiple sources, but the largest is due to the sample's proximity to the superconducting shield. In the presence of a magnetized object or a current, superconductors screen transverse magnetic fields with surface currents via the Meissner effect. If the superconductor's boundary is planar (as is the case in this design), the resultant induced magnetic field is identical to that produced by an identically magnetized object ``mirrored'' over the boundary. This is known as the ``Meissner image'' of the object. The Meissner image of the sample volume is an identically magnetized spheroid, placed equidistant from the superconducting shield but on the other side. Figure \ref{image} shows this mirroring effect. This image spheroid is evidently external to the real sample; therefore its magnetic field is not constant at the location of the real spheroid.\cite{elip95} This field produces a significant non-constant perturbation in the magnetic field internal to the real sample, resulting in large magnetic field gradients. Figure \ref{image} shows the significant variation in the internal field when the image spheroid is introduced. A scheme for canceling these gradients is described in section \ref{gradients}.

\section{Magnetic Gradient Compensation}\label{gradients}
\begin{figure}
\includegraphics[width=0.6\columnwidth]{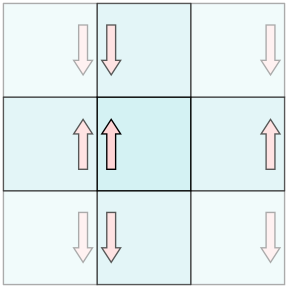}
\caption{\label{infinite} As a result of the ``images of images'' phenomenon, a magnetized object in a cubical superconducting shield experiences a magnetic field that appears as though there is an infinite lattice of such cubes surrounding it.}
\end{figure}
Before it is possible to reduce the magnetic gradients in the sample, the full extent of the problem of magnetic images must be found. The essential problem is that while a planar superconductor acts like a magnetic ``mirror'', a general superconducting enclosure is not so simple. For instance, if a magnetized object is placed between two parallel superconducting planes, one left and one right, it would seem that a magnetic image of the object mirrored across the left plane would cancel the transverse component of the object's field across that plane, and a right-hand image for the right plane. However, the left image's field has a transverse component at the location of the right plane which is not canceled by the right image, and vice versa. Therefore, there must be a ``second order'' image, an image of the left image, mirrored over the right plane, and vice versa, to meet the condition of no transverse field. Indeed, these second order images must have their own third order images, etc. Reassuringly, the distance from the real object to the new pair of images increases each time by twice the distance between the planes $d$, so that the magnetic field between the plates is roughly proportional to $\sum_{n=1}^{\infty} \frac{1}{(2 n d)^3}$, which converges. The resulting layout of images can also be generated by ``mirroring'' the superconducting planes across one another, and allowing the resulting image planes to act as magnetic mirrors themselves. This conceptually cleaner approach allows the quick analysis of the cube shaped superconducting shield in ARIADNE: one simply expects that the field will look like that inside an infinite ``lattice'' of such cubes, each one a mirror image of its neighbors (figure \ref{infinite}).

The benefits of this approach to calculating the magnetic field internal to a superconducting shield are primarily computational. The placement of images can be calculated to a certain order and then truncated, and the magnetic field of this finite number of identical objects will then approximate the magnetic field produced by surface currents in the superconductor. The spheroid is very close to one side of the cubical shield; therefore the field gradients introduced by the images are dominated by the lowest order and only a few orders need be taken to ensure an accurate approximation. Therefore the method allows for rapid magnetic field calculations, and rapid optimization of the gradient reduction system.

The space of potential solutions to the gradient reduction problem is greatly constrained by cost considerations. The short-range nature of the axion coupling necessitates sub-millimeter scale geometry in the design of the $^3$He vessel and magnetic shield; as a result any modification of the shield geometry itself requires precision manufacturing and is prohibitively expensive. Thus the solution is essentially constrained to magnetic sources in the interior of the shield. The simplest configuration of such sources is a single superconducting coil. The axis connecting the centers of the real spheroid and its image is perpendicular to the shield wall and is also an axis of symmetry of the system. Therefore, for a coil to cancel the magnetic field gradients induced by the image it must be located on this axis. There are then only three parameters that can vary: the coil radius, its current, and its location along the axis. A simple algorithm scans the parameter space for the configuration that optimizes some function of the magnetic field across the sample, in progressively finer steps. Two such optimizations were considered: one which minimized the integral of the norm of the magnetic field from the image spheroid across the sample (referred to as field cancellation), and another which minimized the variance of the norm of the total field (referred to as gradient cancellation).

%
\begin{figure}
\includegraphics[width=1.0\columnwidth]{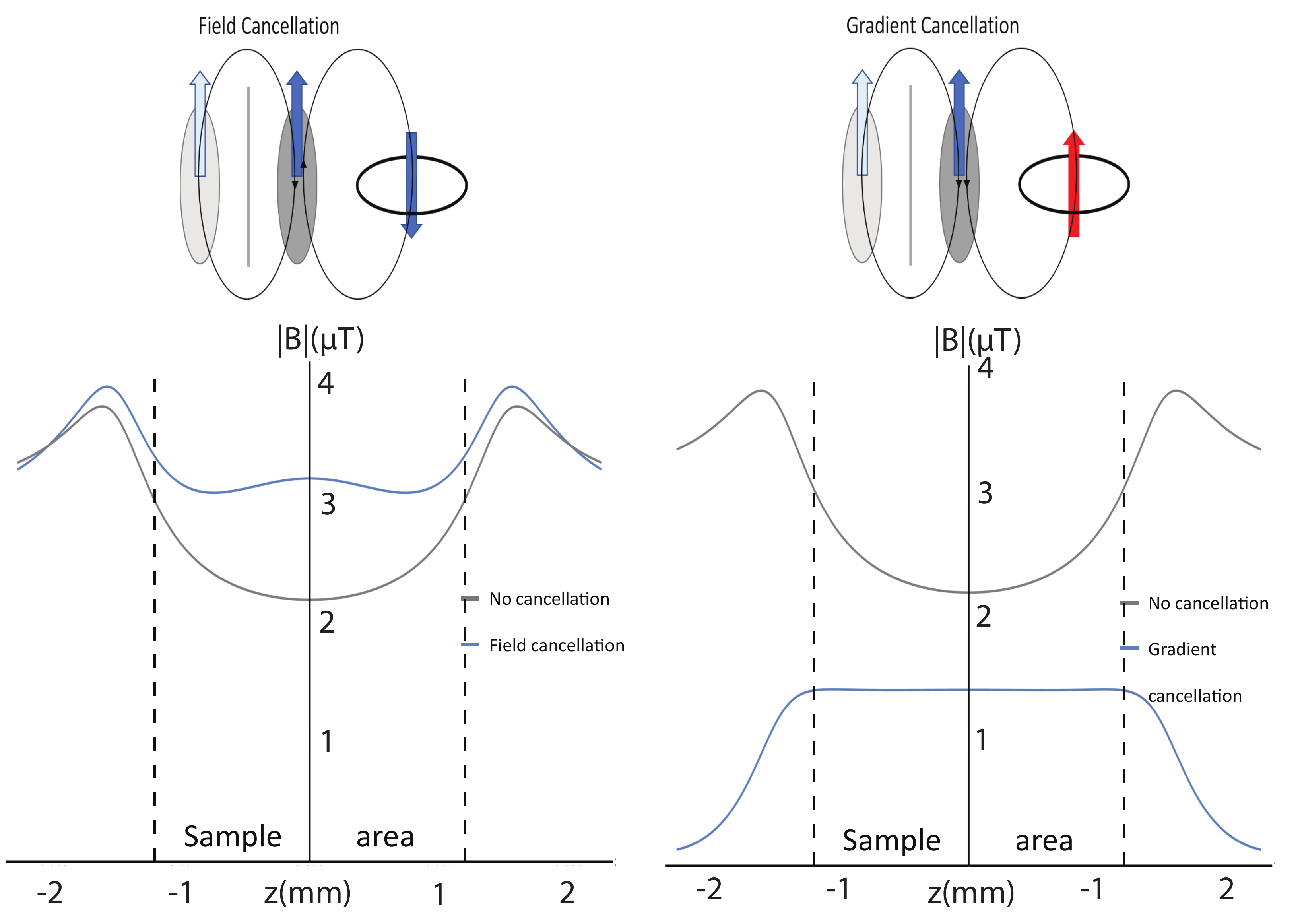}
\caption{\label{gcancel} (left) The field cancellation method seeks to cancel the field from the image spheroid, thereby flattening the total field and allowing it to be tuned to any value. (right) The gradient cancellation approach seeks to ``flatten'' the magnetic field (specifically, minimize the variance of its norm) without regard for its overall value. It reduces the variance by much more than the field cancellation method, but results in a magnetic field which is antiparallel to the spins and nontrivial to tune to an arbitrary value.}
\end{figure}

The main advantage of the field cancellation approach is its versatility. If the field perturbation induced by the image spheroid is cancelled, the resultant total magnetic field is not only flat (in that its norm does not vary) but also unidirectional, along the magnetization axis. This allows the field to be adjusted to any level by a set of modified Helmholtz coils (whose design is described in section \ref{bias}). However, the possible effectiveness of the method is limited, since the perturbed field is bimodal (see figure \ref{image}), unlike the field from a coil. This method can reduce the magnetic field variance by a factor of 3,  using a $3$ mm diameter correction coil carrying a current of $11.6$ mA (figure \ref{gcancel}). 

Rather than attempting to cancel the bimodal perturbed field distribution, the gradient cancellation algorithm finds a different solution. A strong magnetic field opposite the magnetization direction, which falls off gradually at approximately the same rate as the perturbed field, can ``fill in'' the central gap in the field. By this method the magnetic field variance can be reduced by a factor of $10^2$ or greater, depending on manufacturing tolerances, using a $3$ mm coil carrying $1.6$ A of current (figure \ref{gcancel}). The drawbacks of this approach are that the resultant field, though it has an approximately constant norm, does not have a constant direction. Therefore, if one attempts to tune the field value with Helmholtz coils, the result will not necessarily still be flat, as the angle between the flattened field and the tuning field may vary from one part of the spheroid to another. Furthermore, the prevailing direction of the flattened magnetic field is antiparallel to the magnetization axis of the spheroid. This restricts measurement times to the relaxation time T$_1$; however, this time can be made extremely long (on the order of hundreds of hours) for optically pumped helium in Cesium-coated glass containers.\cite{heil95}
Both the field cancellation and the gradient cancellation coils will be installed in ARIADNE's final design.

\section{Bias Field Control}\label{bias}
\begin{figure}
\includegraphics[width=0.6\columnwidth]{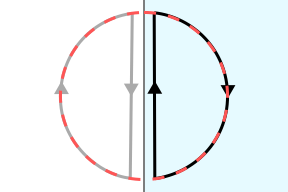}
\caption{\label{dCoils} The ``D coil'' design employs the Meissner image (grey) of the real, ``D'' shaped coil (black) to approximate the field of a Helmholtz coil, represented by the dotted red circle.}
\end{figure}
Conventional nuclear magnetic resonance experiments employ Helmholtz coils to tune the magnetic field felt by the polarized sample, and therefore its Larmor frequency, to a predetermined value. However, since the distance from the $^3$He spheroid to the nearest superconducting boundary is less than the length of its major axis, there is not enough room for a conventional set of Hemlholtz coils on the magnetization axis of the spheroid. Furthermore, any set of coils placed on the magnetization axis will induce magnetic images of themselves across the nearby boundary, potentially inducing further magnetic field gradients in the sample. The novel layout shown in figure \ref{dCoils} solves both problems. A semicircular, or ``D'' shaped coil is placed with its straight side facing the superconducting boundary. The resultant Meissner image completes the outer, circular portion, while its straight side provides an equal but opposing current very nearby that of the real straight portion, approximately cancelling its magnetic field. The sum of the magnetic fields of the real ``D'' shape and its image then approximates that of a circular coil, with the same radius as the semicircle and centered at the superconducting boundary. The approximation improves the closer the straight side of the semicircle is brought to the boundary -- in practice they will be as close as manufacturing tolerances allow. By this approach, large Helmholtz coils can be approximated nearby superconducting boundaries, and since the Meissner image across the nearby boundary forms a part of the Helmholtz field there are no image-induced magnetic field gradients.

By analogy with the ``lattice of cubes'' view of the total magnetic field of a source in a cubical superconducting shield (figure \ref{infinite}), the field of a ``D coil'' inside a cubical shield approximates that of an infinite lattice of rectangular prisms. The prisms have side lengths equal to that of the cube on two sides and twice that of the cube along the axis normal to the side of the cube over which the ``D coil'' is reflected, and each one contains a full circular coil halfway along the long side of the prism, each a mirrored copy of its neighbors. This approximation allows for rapid, analytic calculations of the total field from the ``D coils'' and all their images, and investigation of how those images affect the Helmholtz field. 
Then the parameters of the Helmholtz and cancellation coils can be optimized, after which the optimal configuration is subject to finite element analysis using COMSOL multiphysics software.

One consequence of the lattice of images of the Helmholtz coils is a change in the usual requirement that the distance between the coils is equal to their radius. In the case of ``normal'' Helmholtz coils, the condition ensures that the second directional derivative of the norm of the magnetic field along the axis of the coils vanishes at the center. However, when the images of the coils are included, this derivative becomes nonzero. To again minimize the derivative, the distance between the coils must be slightly increased. The necessary change is greater the larger the coils are, because larger coils are farther apart and therefore closer to the walls of the superconducting shield. For example, $2$ cm coils must be moved apart an extra $0.4$ mm to compensate for the image fields, while $3$ cm coils must be set apart $4$ mm extra, in a box of dimensions 5 cm per side.

\section{Magnetic Shielding Design considerations}\label{design}

\subsection{Quartz block design with integrated sensors and magnetic field control}

\begin{figure}
\includegraphics[width=1.0\columnwidth]{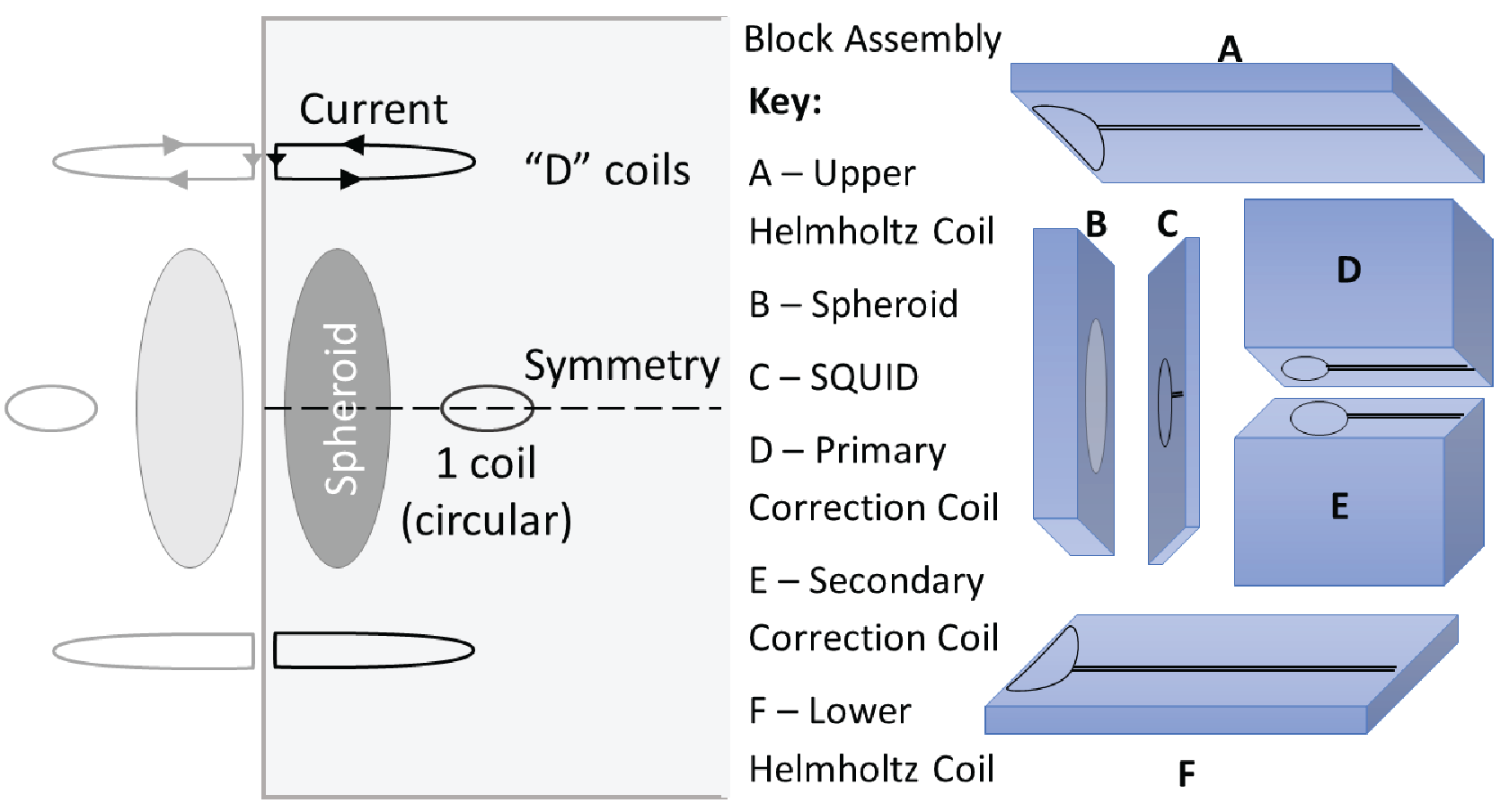}
\caption{\label{construction} Schematic of quartz block construction prior to Nb coating, showing $^3$He sample block as well as patterned coils on additional quartz sections for SQUID readout, magnetic bias field control, magnetic gradient control, and magnetic calibration testing. }
\end{figure}

\begin{figure}
\includegraphics[scale=0.5]{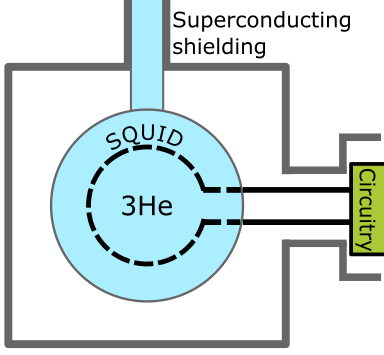}
\caption{\label{shield2} A schematic view of the superconducting shield around the $^3$He sample, showing the necessary holes, which limit the shielding factor.}
\end{figure}

The quartz 3He sample chamber cavity is fabricated by fusing together two pieces of quartz containing hemi-spheroidal cavities. After this, other blocks of quartz with patterned Nb wires are epoxied to the block containing the 3He cell, and the entire assembly will be coated with 1.5 $\mu$m of Nb metal, as shown in Fig. \ref{construction}. However, the SQUID used to measure the magnetic field of the $^3$He must be powered, and its signal read out, by wires. These wires must therefore exit the superconducting shield. Additionally, $^3$He plumbing must pierce the shield so that the hyperpolarized helium can enter the sample volume. There must therefore be holes in the shield (see figure \ref{shield2}), resulting in a reduced shielding factor. COMSOL \cite{comsol} multiphysics software was used to characterize the shielding factor, and to develop an understanding of its dependence on design geometry.

\begin{figure}
\includegraphics[scale=1]{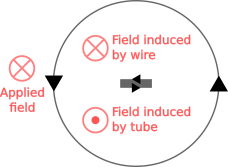}
\caption{\label{coax} A cross-sectional view of the current (black arrows) and magnetic field (red vectors) inside a superconducting tube with a central superconducting wire in an applied magnetic field. Since the shielding from the tube is imperfect, an azimuthal current is induced in the surface of the wire, resulting in a worse shielding factor in the vacuum between the wire and the tube.}
\end{figure}

\begin{figure}
\includegraphics[width=1.0\columnwidth]{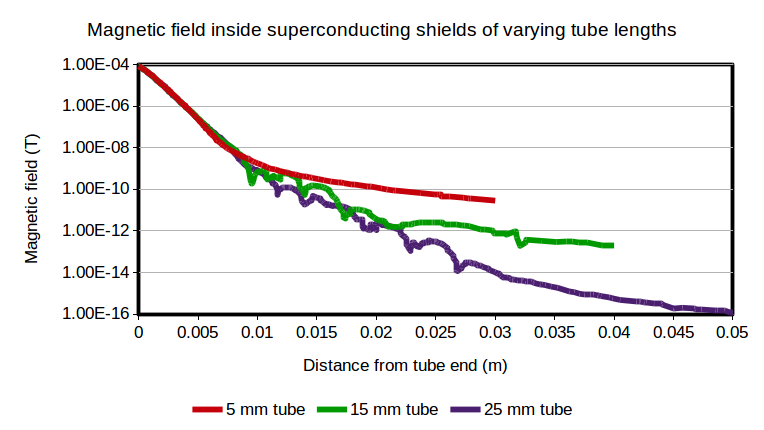}
\caption{\label{shieldLines} Line plots of the magnetic field along the axis of the tube emerging from the superconducting shield, as a function of distance from the tube end. Tube lengths between $5$ mm and $25$ mm are plotted.}
\end{figure}

\begin{figure}
\includegraphics[width=1.03\columnwidth]{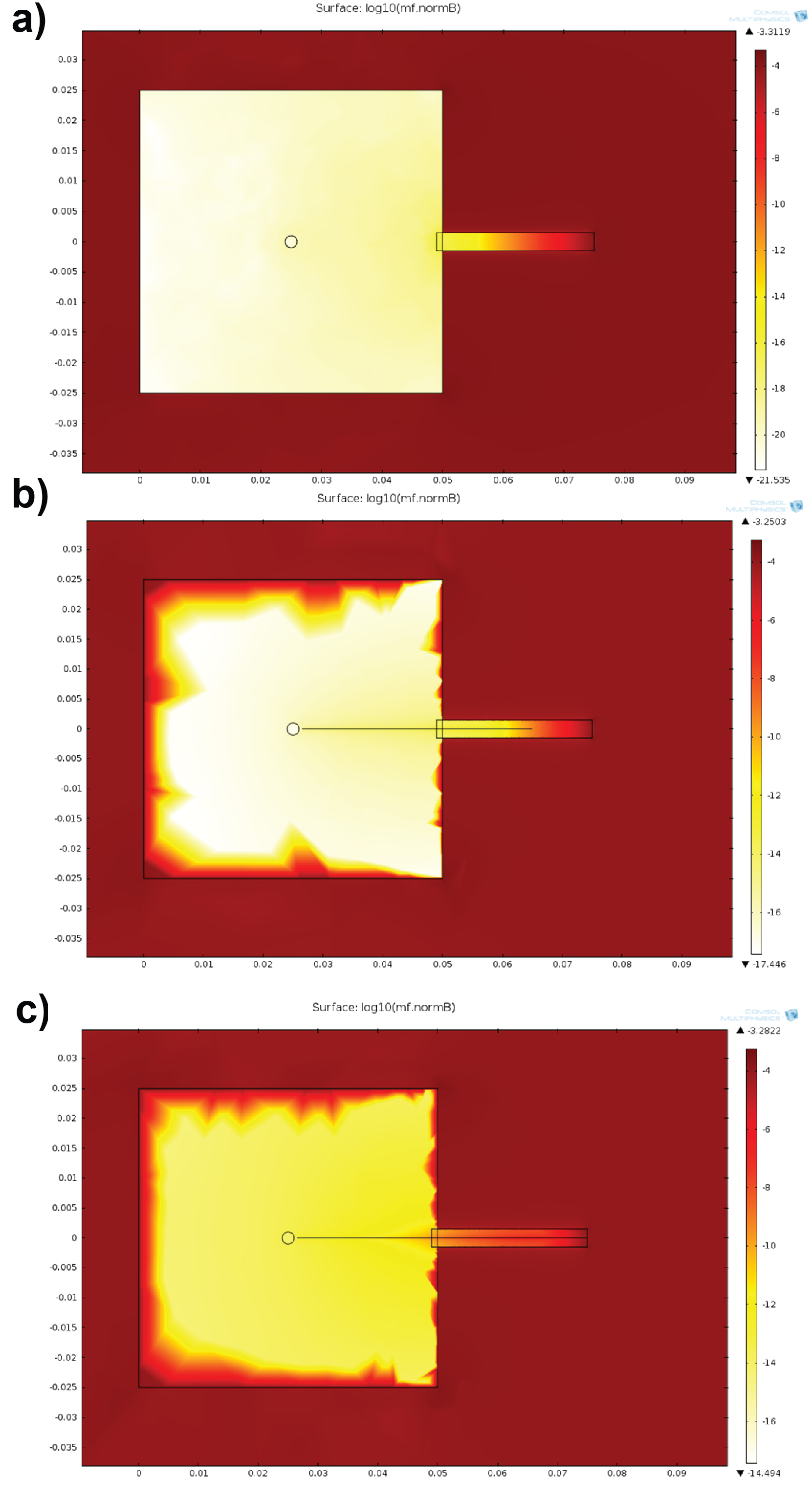}
\caption{\label{coaxLines} Plots of the magnetic field along the axis of the tube emerging from the superconducting shield, where the length of the superconducting portion of the central wire is varied from no superconducting wire, to one that extends to a point $10$ mm from the end of the tube, to one that extends to the end of the tube.}
\end{figure}

Initial COMSOL simulations calculated the shielding factor of a cube with a single uncapped tube extending from one side. It was found that the resultant interior field depended strongly on the angle between the applied exterior field and the axis of the cylinder. This phenomenon is explained physically by the fact that surface currents running in the azimuthal direction around the tube can effectively cancel an axial magnetic field, while there is no complete planar path for current to travel with the axis of the tube in-plane. External fields transverse to the axis of the tube are therefore screened less effectively. For a $15$ mm long tube with a $3$ mm diameter, screening was $10^{15}$ order for a longitudinal magnetic field (along the cylinder axis) and $10^{13}$ for one transverse to the axis. Further simulations calculated the shielding factor of a superconducting cube with two holes, located on adjacent sides ($90$ degrees apart), and found that it was not significantly worse than the transverse field shielding factor of a block with a single hole.

An important consideration in the calculation of the shielding factor of the superconducting enclosure is the effect of the signal wires themselves. There is a topological difference between the cross-section of an empty superconducting tube (where the interior vacuum is simply connected) and that of a tube with a central superconducting wire inside (where the vacuum is not simply connected), whose design mirrors that of a coaxial cable. As a result, even small diameter central wires induce a large change in the shielding factor of the tube. Figure \ref{coax} shows the physical process (i.e. the current density) by which a central superconducting wire reduces the shielding factor. In effect, the residual (non-screened) magnetic field inside the tube induces azimuthal surface currents in the wire. By Lenz's law, these currents reduce the magnetic field inside the wire, but therefore they must increase the magnetic field in the space between the wire and the tube (since the external magnetic field of a finite solenoid is opposite to its internal field).

When the central superconducting wire is included in our simulations, the shielding factor of the single-tube shield drops to $10^6$ order for both a magnetic field parallel to the cylinder axis and for a transverse field. Two approaches are used to improve this shielding factor. Firstly, simply increasing the ratio of the tube's length to its diameter increases the shielding factor. For a tube which is open on both ends, axial magnetic fields are suppressed at the center of the tube exponential in the length-to-diameter ratio \cite{tube76}. For example, our simulations show that increasing the tube length from $5$ mm to $25$ mm results in a factor of $10^5$ improvement in the shielding factor (figure \ref{shieldLines}).

Another method of increasing the shielding factor is to fabricate the central wire from a mix of materials: superconducting metal for the part closest to the SQUID, and ``normal'' (i.e. non-superconducting) metal for the remainder of the length of wire. At the relatively low frequency (~$100$ Hz) of ARIADNE's search, normal metals have little to no magnetic shielding properties, so the shielding factor is not significantly reduced by a central wire made of normal metal. However, Johnson noise has the potential to drown out the small fictitious magnetic field if any normal metal is placed too close to the SQUID and inside the shield. There is therefore an ideal length of superconductor, which balances the reduction of Johnson noise with the maximization of the shielding factor. Figure \ref{coaxLines} shows the magnetic field as a function of distance along the axis of the tube for no superconucting wire, a superconducting wire that extends $15$ mm down the $25$ mm tube, and a superconducting wire that extends to the end of the tube. As expected, the greater the length of superconducting wire present, the less shielding is achieved. However, the difference between the case with $15$ mm of superconducting wire in the tube and that with no superconductor at all is comparatively small, suggesting that reducing the length of superconductor could be an effective strategy to recover high shielding factors.

\begin{figure}
\includegraphics[width=1.0\columnwidth]{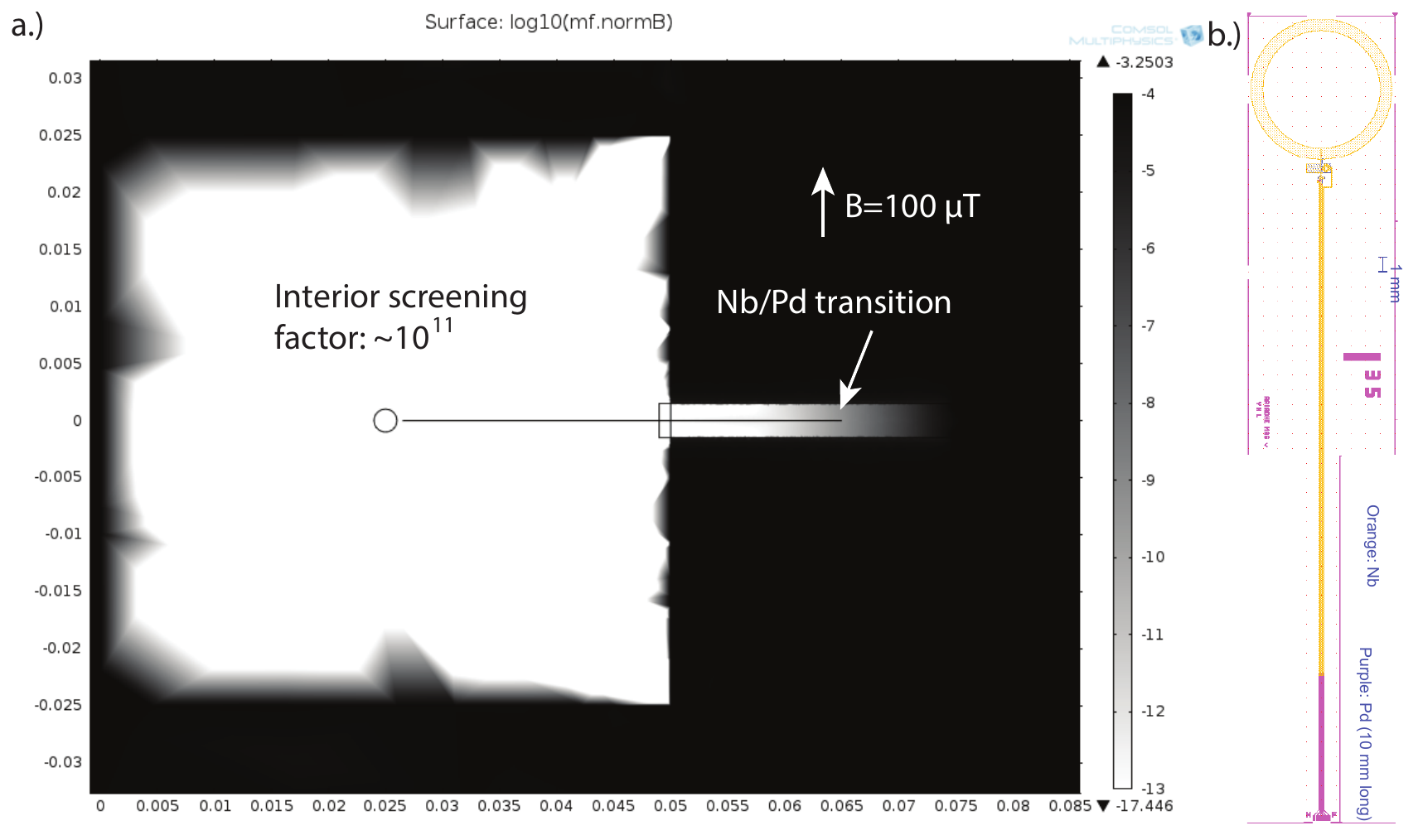}
\caption{\label{shieldFactor} A numerical simulation of the magnetic shielding factor of the pierced shield, including the superconducting SQUID. Right: the design of the SQUID to be used in ARIADNE.}
\end{figure}

The design chosen for ARIADNE is indicated in Fig. \ref{shieldFactor}, where the central conductor changes from superconducting Nb to normal metal Pd 15mm from the block in the 25mm tube, and exhibits a predicted shielding factor of approximately $10^{11}$, which should be approximately 3 orders of magnitude greater than that needed to achieve the design sensitivity.

\section{Conclusion}
We have presented a method to control the magnetic field and gradients within a gaseous sample of polarized nuclear spins in close proximity to a superconducting boundary. We have analyzed shielding solutions using finite element simulation. The approach could be used for other precision NMR or electron-spin resonance (ESR) studies in the neighborhood of a superconducting boundary \cite{quax, casper, quax-gsgp}. 

\section{Acknowledgements}
We thank M. Arvanitaki and C. Lohmeyer for discussions, and H. Mason for early assistance with numerical simulation work.  This work is partially supported by the U.S. National Science foundation, grants PHY-1509805 and PHY-1510484.
\nocite{*}
\bibliography{main.bib}

\end{document}